\documentclass[refree]{raa}
\usepackage{graphicx,times}
\usepackage{natbib}
\usepackage{amssymb,amsmath}
\bibpunct{(}{)}{;}{a}{}{,}

\usepackage[a4paper=true,pagebackref=true]{hyperref}
\hypersetup{pdftitle = The title of my PDF, pdfauthor = My name, pdfsubject= The subject, pdfkeywords = keyword1 keyword2 keyword3}
\hypersetup{colorlinks = true, linkcolor = green, anchorcolor = red, citecolor = blue, filecolor = red, pagecolor = red, urlcolor = red}

\begin{document}

\title{The Stellar System HIP 101227: Is it a Binary, a Triple or a Quadruple System?
}
 \volnopage{Vol.0 (20xx) No.0, 000--000}      
 \setcounter{page}{1}          

\author{Z. T. Yousef,\inst{1}
A. Annuar,\inst{1}\thanks{Corresponding author e-mail: adlyka@ukm.edu.my (A.A)}
A. M. Hussein\inst{2},
H. M. Al-Naimiy\inst{3,4},
M. A. Al-Wardat,\inst{3,4}\thanks{E-mail: malwardat@sharjah.ac.ae}
N. S. A. Hamid\inst{1},
M. F. Talafha\inst{3,4}
}
   \institute{Department of Applied Physics, Faculty of Science and Technology, Universiti Kebangsaan Malaysia, 43600 UKM Bangi, Selangor, Malaysia;\\
        \and
               Department of Physics, Al al-Bayt University, Mafraq 25113, Jordan\\
        \and
        Department of Applied Physics and Astronomy, University of Sharjah, P.O.Box 27272 Sharjah, United Arab Emirates\\
        \and
             Sharjah Academy for Astronomy, Space Sciences and Technology SAASST, University of Sharjah, P.O.Box 27272 Sharjah, United Arab Emirates\\
\vs\no
   {\small Received~~20xx month day; accepted~~20xx~~month day}}

\abstract{
In this paper, we present the analysis of the stellar system HIP 101227 to determine the actual number of components in the system, and their properties. We use dynamical modeling and complex spectrophotometric (involving atmospheric modeling) techniques with recent data, to determine the physical properties and orbital solution for the system, respectively, with better accuracy than past studies. Based on our analysis, we found that the system is more consistent with being a quadruple rather than a binary, or a triple system as that suggested by previous studies. The total mass of the system determined from our SED analysis is 3.42 $\pm$ 0.20 $M_{\odot}$, which are distributed almost equally between the four stars. The stars are found to be zero-age main sequence stars; i.e., at the last stage of pre-main sequence,with age less than 200 Myr and spectral types K0. All four stars have very similar physical characteristics, suggesting that fragmentation process is the most likely theory for the formation and evolution of the system.
}

   \authorrunning{Z. T. Yousef et al. }            
   \titlerunning{HIP 101227: Binary, Triple or Quadruple?}  
   \maketitle
\section{Introduction}
\label{sect:intro}
Most of the stars in our Universe are believed to be part of binary or multiple stellar systems~(\citealt{duquennoy1992binaries}). The study of these objects is crucial for testing the formation and evolutionary models of stars. This is because they can provide us with the direct measurements of stellar masses (e.g., through dynamical modeling), which is the key property that determines the life cycle of stars. The vast majority of these systems discovered so far are close visual binary and multiple stellar systems~(CVBMSs). This means that they appear as a single star even
with the largest ground-based telescopes due to their very close separation; i.e., $<$ 0.5$\arcsec$~(e.g., \citealt{hilditch2001introduction}).
This type of stellar systems can only be identified using high-resolution
observational techniques such as speckle interferometry~(e.g., \citealt{labeyrie1970attainment} and \citealt{balega1977speckle}) and adaptive optics.

One of the most common methods used to study the properties of binary stellar systems
is photo-dynamical modeling. In this technique, the physical and orbital properties of the stars are obtained by modeling the light curves of binary systems during eclipse~(\citealt{borkovits2016comprehensive,borkovits2019photodynamical},~\citealt{koccak2020photometric}, and~\citealt{surgit2020absolute}). This method however, can only be used for eclipsing binary systems, which are relatively rare.

In 2002, \cite{al2002spectral} introduced a novel method to determine the complete physical properties of CVBMSs through spectral energy distributions (SED) modeling; i.e., atmospheric modeling. The data required for this technique are the magnitude differences obtained from speckle interferometry measurements, and color indices. These data are widely accessible, and therefore the technique can be used to study the properties of all types of CVBMSs, including face-on orbit binaries. This method can be used even in the absence of orbital information and spectroscopic data of the systems. These make it an ideal technique to study the properties of these systems. A series of papers that determine the complete set of physical and orbital properties of CVBMSs using this method and dynamical modeling technique developed by \cite{tokovinin1992complementary}, respectively, have been published for many systems (e.g.,
\citealt{al2007model},
\citealt{al2009parameters},
\citealt{al2009parametersb},
\citealt{al2012physical},
\citealt{al2014speckle}, \citealt{al2014physical}, \citealt{al2016physical},
\citealt{masda2016physical},
\citealt{al2017physical}, \citealt{masda2018physical},  \citealt{masda_2019} and  \citealt{masda2019physical}). These analytical papers aim at taking the advantage of CVBMSs in estimating the key properties of the stellar components which is crucial in understanding the formation and evolution of the systems.

In this paper, we will present the SED and orbital modelings of the triple stellar system candidate HIP 101227 (\citealp{tokovinin18}; \citealp{dommanget2002vizier}). This system is part of our study of the brightest ($V \leq$ 10 mag) and closest ($d \leq$ 100 pc) triple stellar system candidates in our Universe. Our aim for this project is to determine the true nature of the systems; i.e., whether or not they are indeed triple systems, and to study their physical and orbital properties. These are important towards having a better census on CVBMSs in general.

\begin{table*}
\caption{Basic properties of HIP 101227.\label{tab:Basic-data-for-HIP101227}}
\centering{}\begin{tabular}{@{}llcl@{}}
\hline
Properties & Parameters             & Value          & Reference \\ \hline
Position   & Right Ascension            & $20^{\rm h}31^{\rm m}07^{\rm s}.77$   &  \href{http://simbad.u-strasbg.fr/simbad}{SIMBAD}         \\
           & Declination        &  $+33{^\circ}32'34''.45$     &  \href{http://simbad.u-strasbg.fr/simbad}{SIMBAD}          \\
Magnitude [mag]  & Visual   Magnitude ($m_{\rm v}$) & 8.34            & \cite{esa1997hipparcos}         \\

           & $B$-$V$ (Johnson)           & $0.88\pm0.01$          &  \cite{esa1997hipparcos} \\
            & Visual Extinction($A_{\rm v}$)  & 3.71      & \cite{2011ApJ...737..103S}          \\
           & $B$ (Tycho)                 & $9.45\pm0.02$          &  \cite{2000AA...355L..27H}  \\
           & $V$ (Tycho)                 &$8.44\pm0.01$          &  \cite{2000AA...355L..27H}   \\
Parallax ($\pi$) [mas] & old                 & $22.38\pm1.16$  &  \cite{esa1997hipparcos}\\
   & new                & $20.47\pm0.95$  & \cite{van2007validation}         \\ \hline
\end{tabular}
\end{table*}

HIP 101227 is located at a right ascension of $20^{\rm h}31^{\rm m}07^{\rm s}.77$,
and declination of $+33{^\circ}32'34''.45$ (\href{http://simbad.u-strasbg.fr/simbad}{SIMBAD} catalog). The parallax of the system obtained from the Hipparcos New Astrometric Catalog~(\citealt{van2007validation}) is 20.47 $\pm$ 0.95~mas, which corresponds to a distance of 48.85 $\pm$ 0.05~pc. In Table~\ref{tab:Basic-data-for-HIP101227}, we present the basic information for the system. Based on the Multiple Star Catalog by \cite{tokovinin18} and Catalog of Components of Double \& Multiple Stars by \cite{dommanget2002vizier}, the system is suggested to be a triple system. However this has not been confirmed, and the system has only been analyzed as a binary in previous literatures.

~\cite{malkov2012dynamical} derived the total mass of the system, assuming that it is a binary, using dynamical, photometric and spectroscopic techniques. The total masses determined from these techniques were $1.03\pm$ 0.31 $\rm M_{\odot}$, 1.82 $\rm M_{\odot}$, and 0.85 $\rm M_{\odot}$, respectively. The orbit of the system has previously been determined by \cite{docobo2006new} using positional measurements obtained from the Fourth Catalog of Interferometric Measurements of Binary
Star.\footnote{The Fourth Catalog of Interferometric Measurements of Binary
Stars can be downloaded at \url{http://www.astro.gsu.edu/wds/int4/int4_20.html}} Since then, there are several new positional measurements for the system. We will use these new data in this paper to obtain a more accurate orbital solution for the system. In addition, we will also try to determine the true nature of the system; i.e., whether it is a binary or multiple system, and measure the physical properties of each star in the system.


\begin{table}
\begin{centering}
\caption{Magnitude difference measurements ($\Delta m$) for HIP 101227 in different filters of the visual band.\label{tab:Magnitude-difference-measurement}}
\par\end{centering}
\centering{}
\begin{tabular}{@{}lccc@{}}
\hline
Filter    & $\Delta m$     & Telescope   diameter & Reference \\
($\lambda/\Delta \lambda$) & & [m] &\\ \hline
511 nm/222 nm & $0.48\pm0.12$           & 0.3                     & \cite{esa1997hipparcos}         \\
503 nm/40 nm & $0.28\pm0.15$     & 3.5                     &  \cite{horch2004speckle}         \\
545 nm/30 nm   & $0.41\pm0.03$ &  6.0                     & \cite{balega2006bull}         \\
541 nm/88   nm & 0.17       &  3.5                     & \cite{horch2008charge}         \\
550 nm/40 nm   & 0.63       &  3.5                     & \cite{horch2009ccd}         \\ \hline
\end{tabular}
\end{table}

\section{METHODS}

\subsection{SED Modeling}

In this section, we describe the synthetic SED modeling of HIP 101227 system using the \cite{al2002spectral} technique in order to estimate the physical properties of each star in the system; i.e., the effective temperatures ($T_{\rm eff}$), radii ($R$), gravitational accelerations (log $g$), luminosities ($L$), masses, ages and spectral types ($S_{\rm p}$). We performed the SED modeling by first assuming that the system is a binary, and then a multiple system.

In order to build individual SED models for the system's components, we first calculated the average value for the magnitude difference between the two main components of the system ($\Delta m$) in the $V$-band. All of the magnitude difference measurements for HIP 101227 in the $V$-band are tabulated in Table~\ref{tab:Magnitude-difference-measurement}. The average value calculated from these data is $\Delta m\ = 0.39$ $\pm0.16$.

We then used this value along with the total visual magnitude of the system; i.e., $m_{\rm v}=8.34$ \citep{esa1997hipparcos}, in the following equations to get the apparent magnitudes for the two main components of the system; i.e., component $A$ and $B$:

\begin{eqnarray}
m_{v}^{A} & = & m_{v}+2.5\ {\rm{log}}\ \left(1+10^{-0.4\Delta m}\right)\label{eq:1}\\
m_{v}^{B} & = &m_{v}^{A}+\Delta m\label{eq:2}\\
M_{V} & = & m_{v}+5-5\ {\rm{log}}\ d-A_{V}\label{eq:3}
\end{eqnarray}
where the interstellar extinction $A_{V} =$ 0.0128, calculated using $E(B-V)=0.004$ \citep{lallement2018three}.
We used the apparent individual magnitude values calculated from these equations, along with  $T_{\rm eff}$, mass, $S_{\rm p}$ and bolometric correction information obtained from \cite{gray2005observation} and \cite{lang1992astrophysical}, to calculate the bolometric magnitude $M_{\rm bol}$, $R$, log $g$ and $L$ of the two main components using the well-known equations for main sequence star:
\begin{eqnarray}
M_{\rm bol} & = &M^{\odot}_{\rm bol}-2.5\ {\rm{log}}\left(\frac{L}{L_{\odot}}\right)\label{eq:4}\\
{\rm{log}}\left(\frac{R}{R_{\odot}}\right) & = & 0.5\ {\rm{log}}\left(\frac{L}{L_{\odot}}\right)-2\ {\rm{log}}\left(\frac{T_{\rm eff}}{T_{\rm eff_{\odot}}}\right)\label{eq:5}\\
{\rm{log}}\ g & = & {\rm{log}}\left(\frac{\rm M}{\rm M_{\odot}}\right)-2\ {\rm{log}}\left(\frac{R}{R_{\odot}}\right)+4.43\label{eq:6}
\end{eqnarray}

Next, we used these estimated values as preliminary input parameters for the grids of
Kurucz's line-blanketed plane-parallel models (ATLAS9) to obtain preliminary synthetic SED for stars $A$ and $B$. These were then combined to get the total synthetic SED of the system according to the following equation \citep{al2012physical}:

\begin{eqnarray}
F_{\lambda} \cdot d^{2}  =  H_{\lambda}^{A} \cdot R_{A}^{2}+H_{\lambda}^{B} \cdot R_{B}^{2}
\label{eq:6}
\end{eqnarray}
which can be written as:
\begin{eqnarray}
F_{\lambda}=  (R_{A} /{d} )^{2}  (H_{\lambda}^{A} + H_{\lambda}^{B} (R_{A} /{R_{B}})^{2})
\label{eq:7}
\end{eqnarray}
where $H_{\lambda}^{A}$ and $H_{\lambda}^{B}$ represent the flux from a unit surface
of the system's components $A$ and $B$, respectively, and $F_{\lambda}$
represents the total SED of the entire system.



\begin{figure}
\begin{centering}
\includegraphics[scale=0.50]{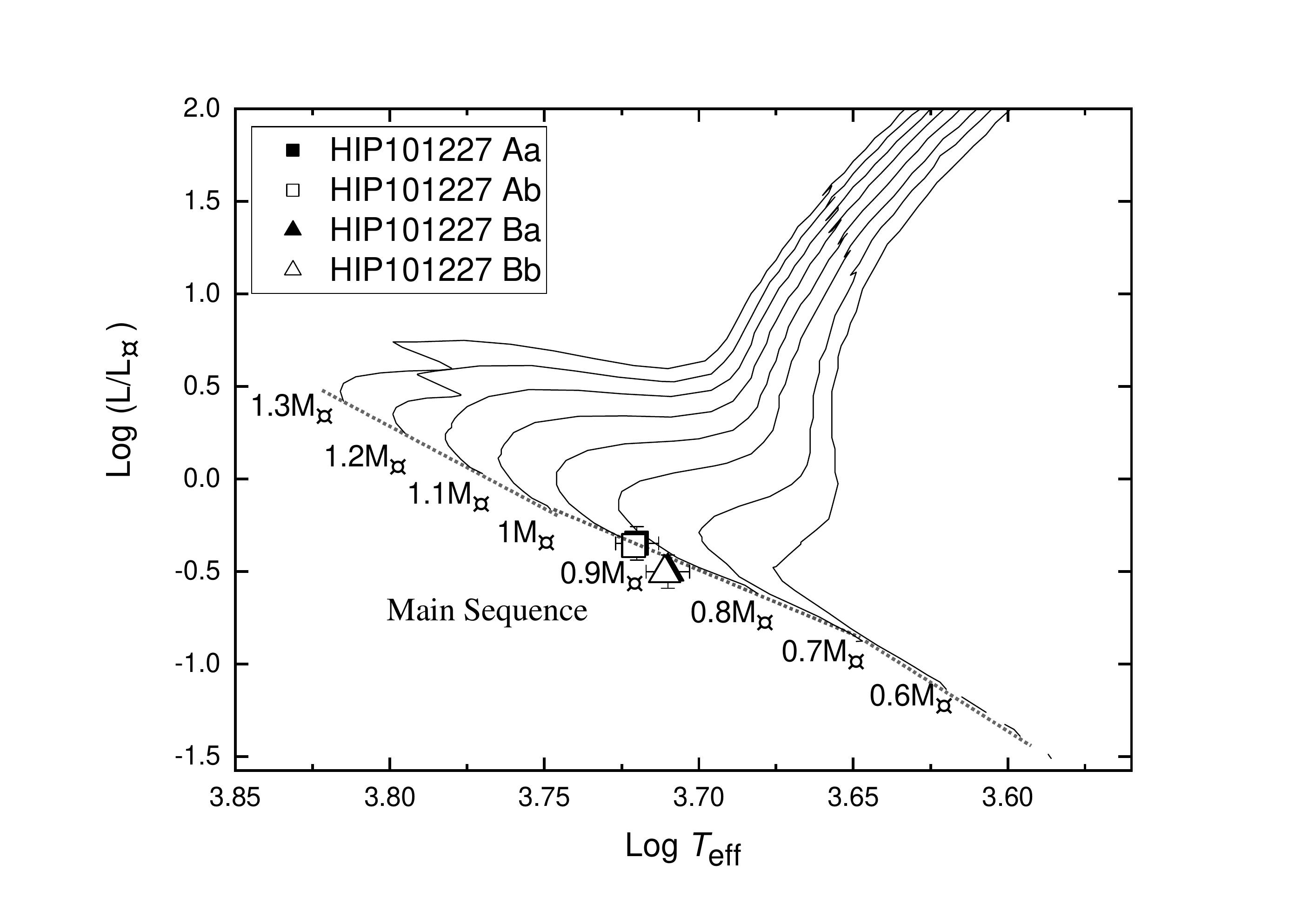}
\caption{The positions of each star in the HIP 101227 system on the evolutionary tracks derived by \cite{girardi2000evolutionary} assuming that it is a quadruple. \label{fig:evolutionary-quad}}
\par\end{centering}
\end{figure}
In order to get the best synthetic SED that will give the most accurate physical properties for the system, we need to perform the above steps in iterative manner using different sets of preliminary input parameters. This should be performed until we obtained synthetic magnitude and color index values that are consistent with that obtained from observations (\href{http://simbad.u-strasbg.fr/simbad}{SIMBAD} catalog) within the error values.

The synthetic magnitudes can be calculated from the synthetic SED using
the following relationships~(\citealt{al2002spectral,al2008synthetic,al2012physical} and references therein):
\begin{eqnarray}
m_{p}=-2.5\ log\frac{\smallint P_{p}(\lambda)F_{\lambda,s}(\lambda) \lambda d\lambda}{\smallint P_{p}(\lambda)F_{\lambda,r}(\lambda) \lambda d\lambda}+ZP_{p}
\end{eqnarray}
where $m_{p}$ represents the synthetic magnitude of the pass-band
$p$, $P_{p}$ is the dimensionless sensitivity function of the pass-band
$p$, $F_{\lambda,s}(\lambda)$ and $F_{\lambda,r}(\lambda)$ are the synthetic
SEDs of the star being studied and the reference star (Vega in this case), respectively. The zero points, $Z$, were taken from \cite{apellaniz2007future}.

\begin{figure}
\begin{centering}
\includegraphics[scale=0.55]{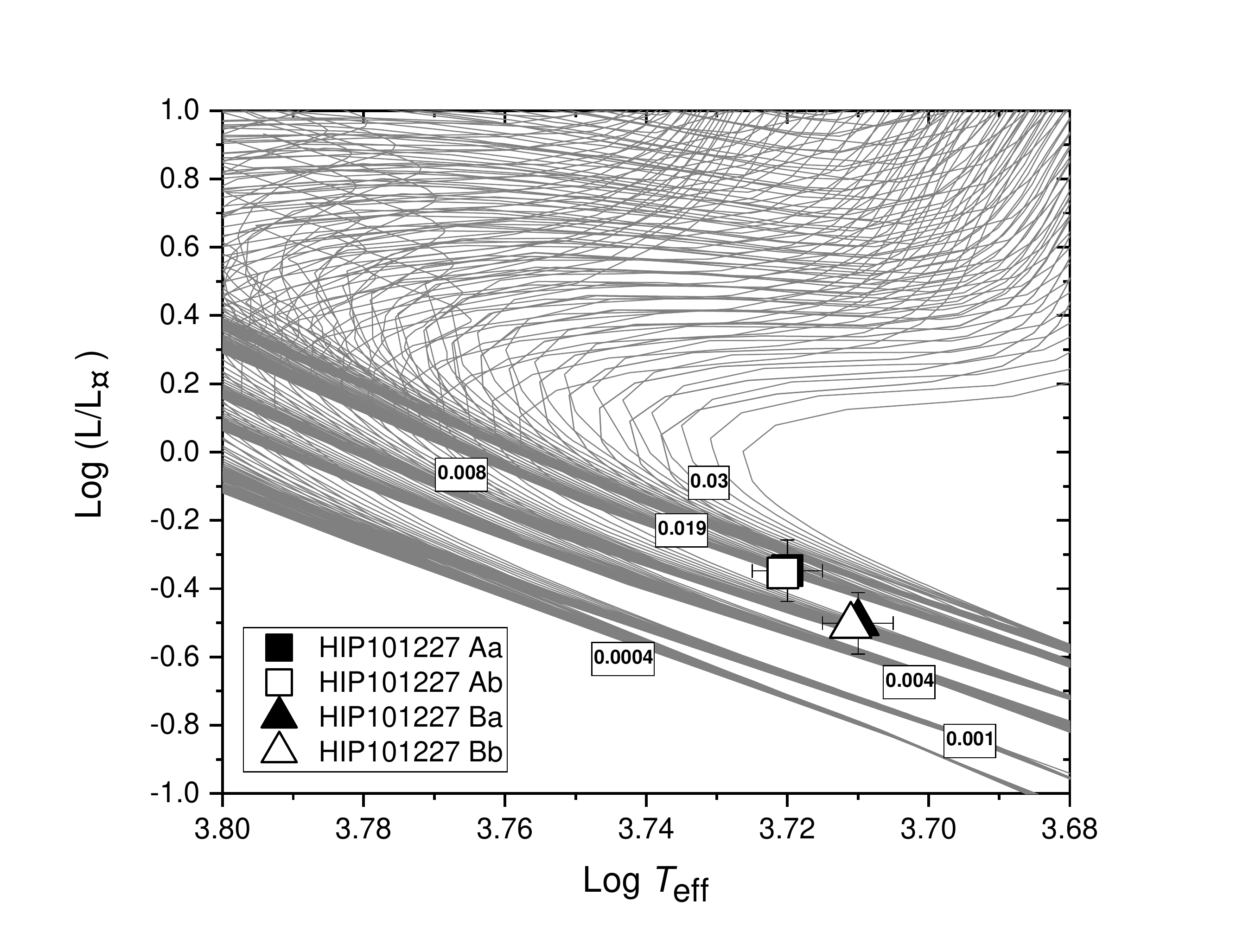}
\caption{The positions of each star in the HIP 101227 system on the isochrones for low and intermediate-mass stars with different metallicities derived by \cite{girardi2000evolutionary} assuming that it is a quadruple.}
\label{fig:iso-quad}
\par\end{centering}
\end{figure}

\subsection{Orbital Analysis}

 The orbital properties of HIP 101227 were determined using the dynamical modeling technique developed by \cite{tokovinin2016orbits} (the original paper is \cite{tokovinin1992complementary}). This method uses the least-square fits with weights inversely proportional to the observational errors to give the final orbital parameters with their errors using the IDL code ORBIT  \citep{tokovinin2016orbits}{\footnote{The IDL code ORBIT can be downloaded at http://www.ctio.noao.edu/{$\sim$}atokovin/orbit/}}. We used relative position measurements from speckle interferometric observations to determine the period ($P$), epoch of passage through periastron ($T_{\rm o}$), eccentricity ($e$), semi-major axis ($a$), inclination ($i$), longitude of the periastron ($\omega$), and the position angle of the line of nodes ($\Omega$) for the orbit of the two main components of the system.

The code requires the preliminary values of $P$, $T_{\rm o}$, $e$, $a$, $i$, $\omega$, $\Omega$, the positional measurements, and radial velocities of the system. We obtained positional measurements for the system from the Fourth Catalog of Interferometric Measurements of Binary Stars$^{1}$ and \cite{2018AJ....155..215M}. The latest measurements, which were not available when \cite{docobo2006new} did their orbital analysis of the system, are listed in Table~\ref{tab:The-latest-relative-position}. We obtained the radial velocities of the whole system from \cite{2012yCat..35470013T} and \cite{2006AstL...32..759G}; i.e., $-23.5\pm 0.04$ km s$^{-1}$ and $-25.6\pm 0.90$ km s$^{-1}$, respectively. The radial velocities for the main components of the system, $A$ and $B$, were obtained from \cite{2018A&A...612A..96F}; i.e., $-31.25 \pm0.15$ km s$^{-1}$ and $-16.88\pm 0.15$ km s$^{-1}$, respectively. We input these values in the ORBIT code to determine the best-fit orbit for the system.

\begin{table*}
\begin{centering}
\caption{The latest relative positional measurements for the HIP 101227 system (that was not included in \citealp{docobo2006new}) taken from the Fourth Catalog of Interferometric Measurements of Binary
Stars. For all positional measurements, please refer to the catalog.\label{tab:The-latest-relative-position}}
\par\end{centering}
\centering{}\begin{tabular}{@{}lccccc@{}}
\hline
Epoch     & Theta ($\theta^{\circ}$) & Rho ($\rho$) & $\lambda$ (nm) & Telescope aperture (m) & Reference \\ \hline
2001.7635 &   139.4& 0.096& 541 &  3.5&   \cite{2008AJ....136..312H}\\
2002.7986&  153.8 &  0.080& 600& 6.0 &
\cite{2013AstBu..68...53B}\\
2005.5184 &   60.9 &   0.085& 520 & 3.5 &
\cite{2008AJ....135.1803D}\\
2006.5223 & 75.5      & 0.099    & 754   & 3.5                & \cite{2008AJ....136..312H}         \\
2006.5223 & 79.1      & 0.112    & 754   & 3.5  & \cite{2008AJ....136..312H}  \\
 2007.6018 & 266.0& 0.144& 550 & 4.0 &\cite{2018AJ....155..215M}\\
2007.8197 & 89.9     & 0.149    & 550   & 3.5                & \cite{2010AJ....139..205H} \\
2008.4563&273.4 &0.166 & 550 & 4.0 &\cite{2018AJ....155..215M}\\
2008.546  & 97.4      & 0.203    & 530   & 0.7                & \cite{gili2012relative}         \\\hline
\noalign{\smallskip}
\end{tabular}
\end{table*}

\section{RESULTS AND DISCUSSION}

\subsection{Binary, Triple, or Quadruple?}
Based on the physical properties that were obtained from the best-fit SED of the system as a binary (Table \ref{tab:binary-parameter} of the Appendix), we plot the stars on the stellar evolutionary tracks diagram by \cite{girardi2000evolutionary}. This is shown in Figure~\ref{fig:evolutionary} (see Appendix). The figure shows that star $A$ is more evolved than star $B$ even though they have similar masses (mass$_{A}$ $=$ 0.79 $\pm$ 0.20 $M_{\odot}$ and mass$_{B}$ $=$ 0.75 $\pm$ 0.20 $M_{\odot}$). We also plot the stars on the metallicity isochrones by \cite{girardi2000evolutionary} (Figure~\ref{fig:iso-binary} of the Appendix). Based on the figure, we found that both stars deviate significantly from the isochrones tracks. These results provide evidence that HIP 101227 is likely not a binary system.

Therefore, we proceeded our analysis by assuming that the system is now a triple system as that suggested by \cite{tokovinin18} and \cite{dommanget2002vizier}. Based on our analysis of the system as a binary, we found that star $A$ has a larger radius than star $B$, thus has a higher probability of consisting of two stars. Hence, we further analyzed star $A$ as a sub-binary system consisting of star $A_{\rm a}$ and $A_{\rm b}$, assuming that they have very similar properties (i.e., $\Delta m =$ 0). The synthetic SED for star $A$ represents the total SED for both stars. We repeated our analysis as described earlier in order to build the individual SEDs for the three stellar components, and measure their physical properties.

We then used the results that we obtained (Table \ref{tab:triple-A-parameter} of the Appendix) to plot the stars on the stellar evolutionary and metallicity isochrones diagrams (Figure \ref{fig:evolutionarytriple} and \ref{fig:iso-triple} of the Appendix, respectively) by \cite{girardi2000evolutionary}. Based on Figure \ref{fig:iso-triple}, it can be seen that star $B$ is located away from the isochrones tracks. The positions of the stars in the stellar evolutionary diagram are also not consistent with that expected (stars $A_{\rm a}$ and $A_{\rm b}$ are more evolved than star $B$ even though they have similar masses; Mass$_{A_{a}}$ $=$ Mass$_{A_{b}}$ $=$ 0.90 $\pm$ 0.20 $M_{\odot}$ and Mass$_{B}$ $=$ 0.75 $\pm$ 0.20 $M_{\odot}$). Repeating the same analysis by instead assuming that star $A$ is a single star, and star $B$ is a sub-binary system, we obtained similar results (Table \ref{tab:TRIPLE-B-parameter} of the Appendix). These indicate that HIP 101227 is likely not a triple stellar system as well, as that suggested by \cite{tokovinin18} and \cite{dommanget2002vizier}.

Next, we proceeded by assuming that the system is a quadruple consisting of stars $A_{\rm a}$, $A_{\rm b}$, $B_{\rm a}$ and $B_{\rm b}$. Based on our results, we found that the stars fit very well with the metallicity isochrones tracks shown in Figure~\ref{fig:iso-quad}. The properties of the stars are also consistent with the stellar evolutionary tracks shown in Figure \ref{fig:evolutionary-quad}. We therefore conclude that HIP 101227 is a quadruple stellar system consisting of stars $A_{\rm a}$, $A_{\rm b}$, $B_{\rm a}$ and $B_{\rm b}$. In Table~\ref{tab:Final-binary-parameter}, we present the physical properties of the system as measured from the best-fit SED of the system. The synthetic magnitudes and color indices determined from the best-fit SED is tabulated in Table~\ref{tab:Magnitude-difference}, in comparison with the observed values obtained from the \href{http://simbad.u-strasbg.fr/simbad}{SIMBAD} catalog. As can be seen from the table, the synthetic values that we obtained are highly consistent with those measured from observations. We note that there are no significant differences between the total synthetic SEDs that we obtained when assuming that the system is a binary, triple or quadruple system.

  \begin{table}
	\caption{Physical properties of each star in the HIP 101227 quadruple system}.\label{tab:Final-binary-parameter}
	\centering{}\begin{tabular}{@{}lcccc@{}}
		\hline
	Parameters     & Comp. $A_{\rm a}$ & Comp. $A_{\rm b}$ & Comp. $B_{\rm a}$ & Comp. $B_{\rm b}$ \	\\ \hline
$T_{\rm eff} $ (K) & $5250\pm30$    & $5250\pm30$     & $5170\pm20$   & $5170\pm20$                       \\
R ($R_{\odot}$) & $0.81\pm0.02$      & $0.81\pm0.02$    &$0.70\pm0.01$   & $0.70\pm0.01$              \\

 Log $g$   & $4.60\pm0.05$& $4.60\pm0.05$& $4.60\pm0.05$ & $4.60\pm0.05$\\
L ($L_{\odot}$) & $0.45\pm0.05$     & $0.45\pm0.05$   & $ 0.31\pm0.05$   & $ 0.31\pm0.05$                 \\
Mass ($M_{\odot}$) & $0.90\pm0.20$     & $0.90\pm0.20$   & $ 0.81\pm0.20$   & $ 0.81\pm0.20$ \\\hline
	\end{tabular}
\end{table}

Figure~\ref{fig:iso-quad} indicates that stars $A_{\rm a}$ and $A_{\rm b}$ have a metallicity, $Z =$ 0.019, while star $B_{\rm a}$ and $B_{\rm b}$ have $Z =$ 0.008. The total mass of the system is $3.42\pm 0.20\  M_\odot$. This is distributed almost equally between the four stars. Based on the temperatures of the stars, they can be classified as type K0 stars. The very similar physical characteristics of the stars in the system, as presented in Table~\ref{tab:Final-binary-parameter}, suggest that fragmentation process is the most probable theory for the formation of the system as opposed to capture theory. The latter would usually form a system with relatively different properties; e.g., the masses of the stars would be significantly different. Hierarchical fragmentation during rotational collapse has been suggested to produce binaries and multiple systems \citep{zinnecker2001binary}. This mechanism is possible if the spinning disk around an incipient central proto-star is fragmented, as long as it continues to infall \citep{bonnell1994massive}.

  \begin{table}
\caption{Comparison between the total magnitudes determined from the best-fit synthetic SED and that obtained from observations (\href{http://simbad.u-strasbg.fr/simbad}{SIMBAD} catalog).\label{tab:Magnitude-difference}}
\centering{}\begin{tabular}{@{}lccc@{}}
\hline
System                             & Filter & \href{http://simbad.u-strasbg.fr/simbad}{SIMBAD} data base & This work \\ \hline
{Johnson-Cousins} & B      & 9.22            & $9.22\pm1.19$               \\
                                   & V      & 8.34            & $8.34\pm1.08$              \\
                                   & B-V    & 0.88            & $0.88\pm0.11$              \\
{Tycho}             & $B_T$     & $9.45$             & $9.46\pm1.23$              \\
                                   & $V_T$     & 8.44             & $8.43\pm1.09$              \\
                                   & $B_T-V_T$  & 1.01            & $1.01\pm0.13$              \\
$\Delta m$                                 & $\cdots$      & 0.39            & $0.39\pm0.05$              \\ \hline
\end{tabular}
\end{table}

We can estimate the age of the quadruple system from the isochrones diagram in Figure~\ref{fig:iso-quad} as follow:
 \begin{eqnarray}
 \label{eq31}
 t_{\rm ms}=f_{\rm ASITR}(m_{A,B}-5~{\rm log}~d+5-A, Z)
 \end{eqnarray}
where $f_{\rm ASITR}$ is the age-synthetic isochrones track function and $Z$ is the metallicity of the star. Based on this,we found that the age of the system is less than 200 Myr, which means that the four stars are hierarchical zero age main sequence stars (at the last stage of the pre-main sequence). This can also be clearly seen from the positions of the stars on the stellar evolutionary diagram in Figure~\ref{fig:evolutionary-quad}.

\subsubsection{Comparison of the Best-fit Synthetic SED with Observational SED}

\begin{figure}
\begin{centering}
\includegraphics[scale=0.50]{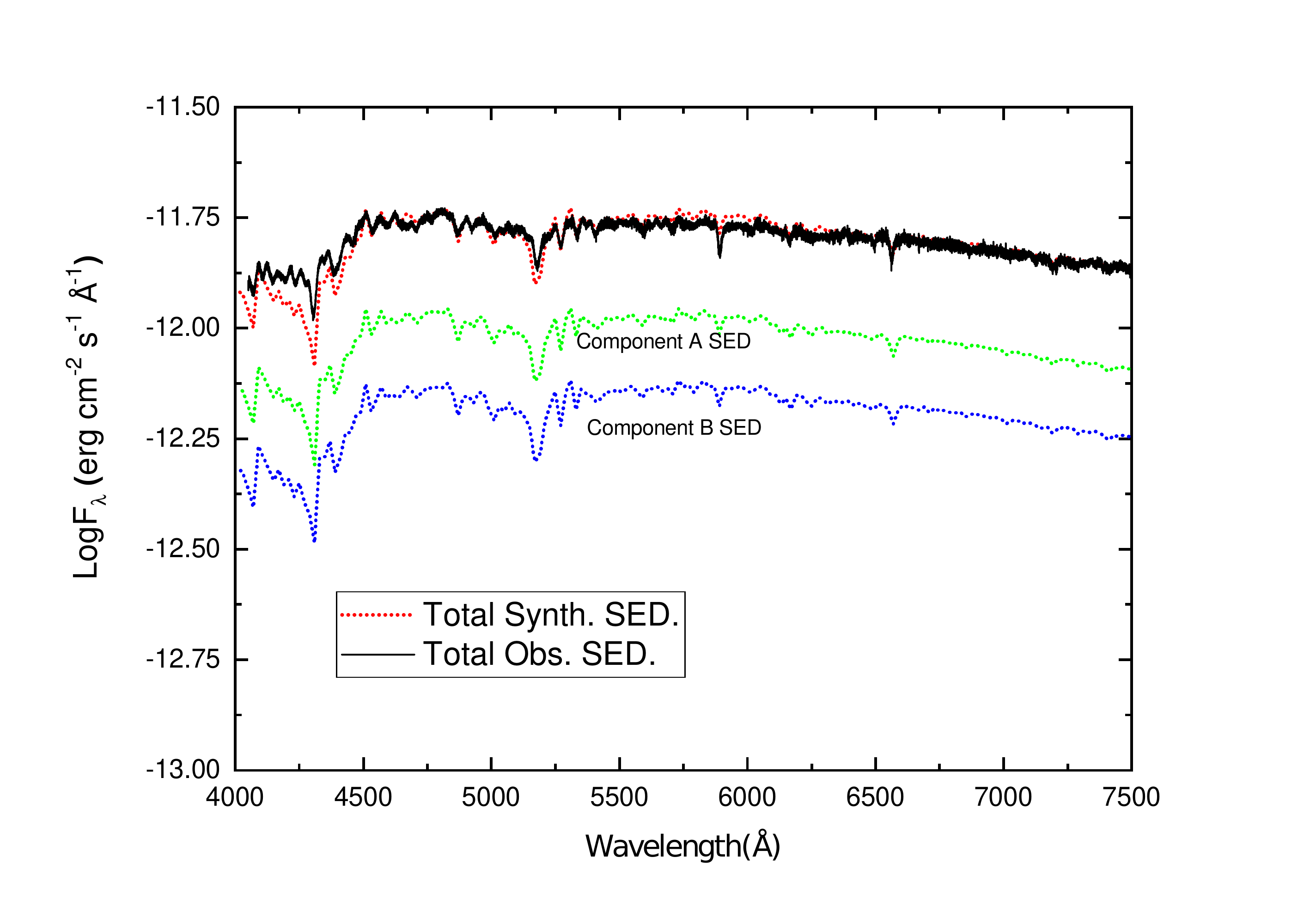}
\includegraphics[scale=0.50]{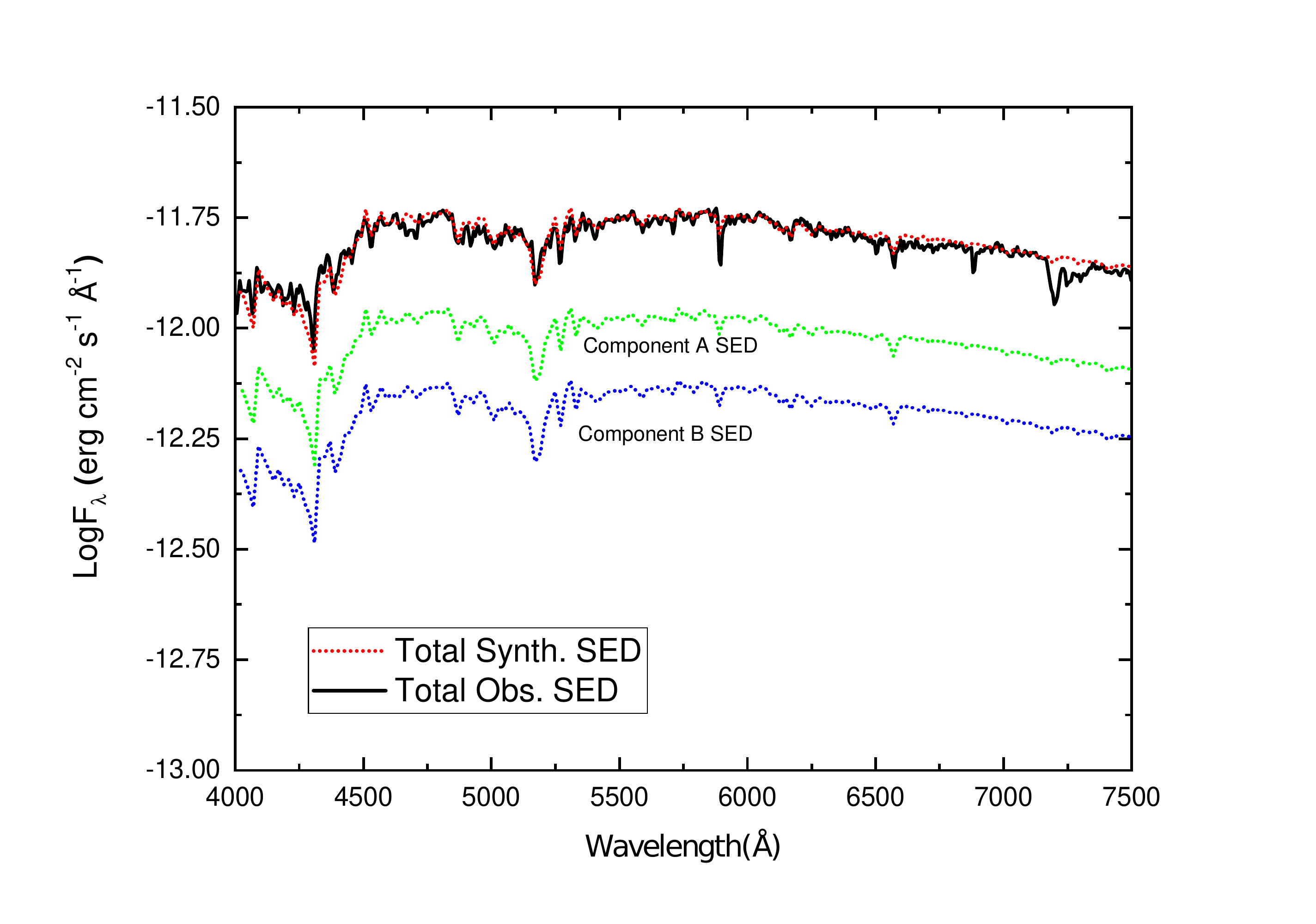}
\caption{The best-fit synthetic SEDs for the main stellar components $A$ and $B$, and the total HIP 101227 system (dashed lines) from our work, in comparison with its total SED obtained from our new observations (solid lines) at Sharjah Observatory (top) and ~\cite{al2002spectral} (bottom).}
\label{fig:Best-Fit-SED}
\par\end{centering}
\end{figure}

In order to test the reliability of our results, which is one of the techniques of Al-Wardat's method, we compare the best-fit total synthetic SED that we determined for the system with its observational SEDs that we obtained specifically for this work, and that firstly obtained by ~\cite{al2002spectral}.

We acquired new observational SED for the system at optical wavelength from the Sharjah Observatory located at Sharjah Academy for Astronomy and Space Sciences in the United Arab Emirates on Julian date 2457201.38422 (one night). Basic  Echelle  Spectrograph (BACHES) was used with SBIG ST-8300 camera at the Plane Wave Corrected Dall-Kirkham (CDK) 17-inch reflector telescope.\footnote{Further details on BACHES can be found at \url{https://academic.oup.com/mnras/article/443/1/158/1481885}} Around 10 spectra were obtained with different exposure times (between 100 s and 300 s). The telescope has a Broad-Band Anti-Reflection (BBAR) multicoated 17$"$~(431.8 mm) aperture, a focal length of 2939~mm~(f/6.8) with a Carbon fibre truss design to minimize focus shift with a collimator focal ratio of f/10. The BACHES spectrograph
has a spectral range of 392~nm to 800~nm. A 50~$\mu$m slit was used, which gives an average spectral resolving power ($R=\lambda/\Delta\lambda$) of 10000. Thorium-Argon
cathode and Halogen lamp with blue filter
was used for calibration and flat-field, respectively. Vega
was used as a reference star.

We stacked the spectra together, and the resultant observational SED for the system is shown in Figure~\ref{fig:Best-Fit-SED}. This is compared with the best-fit synthetic SED that we obtained earlier. We also do the same comparison with the old observational SED obtained from~\cite{al2002spectral} in Figure~\ref{fig:Best-Fit-SED}. Based on this figure, we can see that the del{best-fit }synthetic SED profiles and continuum that we built for the entire HIP 101227 system are generally consistent with both observational SED’s. There is some disagreement between $\sim$4000-4300$\AA$ in our observation. This can be explained by CCD's sensitivity difference between the red and the blue parts of the spectrum. Overall however, these results provide further support for the results that we obtained from our analysis, and the reliability of the \cite{al2002spectral} method in determining the physical properties of CVBMSs.

\subsection{Orbital Properties}

\begin{figure}
\begin{centering}
	\includegraphics[scale=0.55]{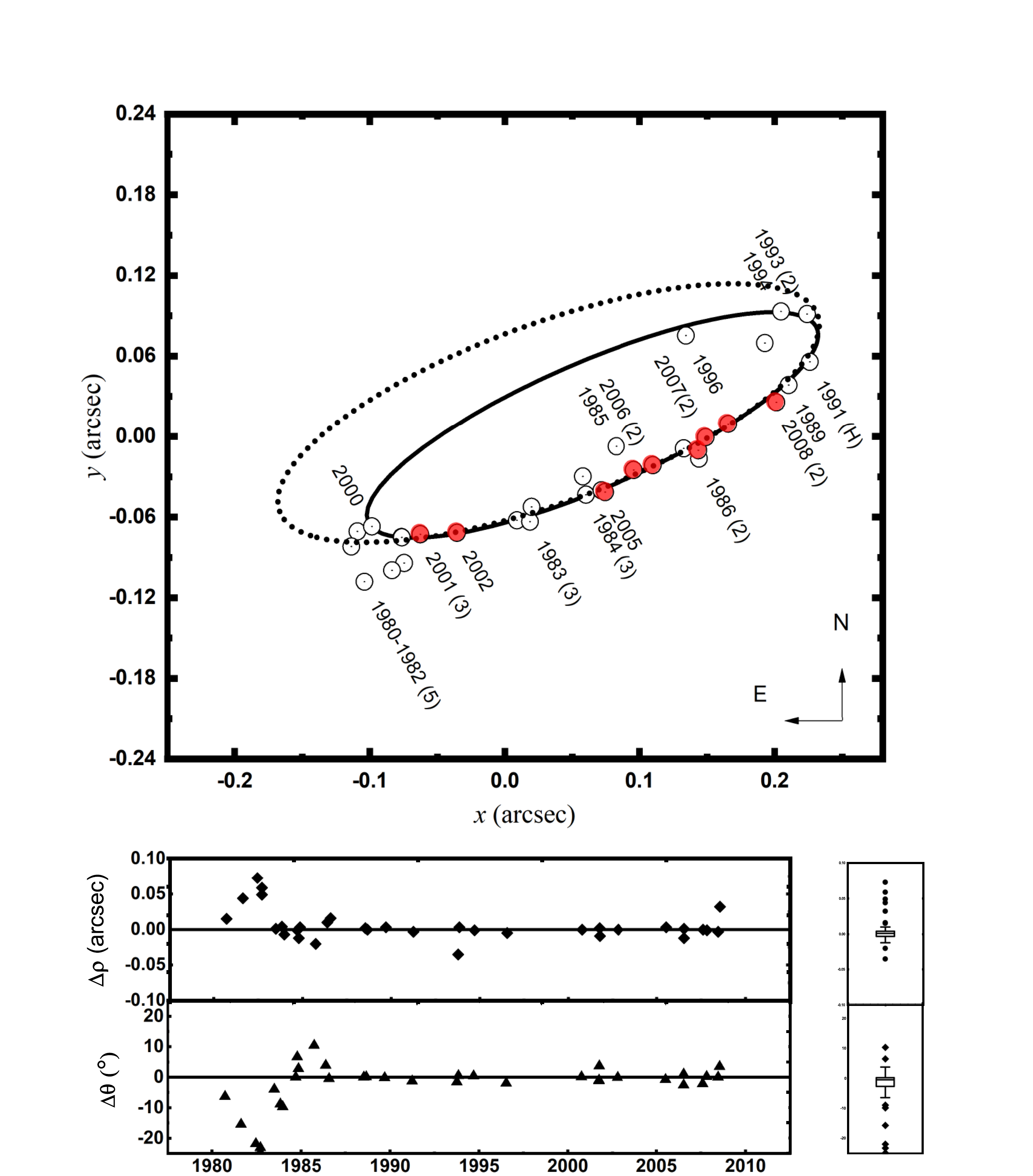}
    	\caption{The orbital solution for HIP 101227 system as determined from our analysis (solid line) and~\cite{docobo2006new} (dotted line). The white and red circles are the old and new positional points (listed in Table \ref{tab:The-latest-relative-position}), respectively. The bottom left panel show the fit residuals, showing the difference between the observed and model values for the angular separation ($\Delta \rho $) and position angle ($\Delta \theta$) of the orbit. The bottom right panel represents the distribution of data based on a five-number summary (“minimum,” first quartile (Q1); median, third quartile (Q3); and “maximum”) and outliers.}
	\label{orbit}
	\par\end{centering}
  \end{figure}

\begin{table}
	\begin{centering}
		\caption{Orbital parameters of HIP 101227 as measured from our analysis in comparison with those obtained by ~\cite{docobo2006new}.\label{tab:The-adopted-orbital}}
		\par\end{centering}
	\centering{}\begin{tabular}{@{}lcc@{}}
	\hline
	 Parameter	& This work      & ~\cite{docobo2006new}\\ \hline
		$P$ (yr.)       & $20.514\pm0.092$                                   & 33.13                                       \\
		$T_{\rm o}$ (yr.)      & $1999.097\pm0.178$                                    & 1974.57                                     \\
		$e$            & $0.5193\pm0.0138$                                          & 0.19              \\
		$a$ (arcsec)    & $0.1989\pm0.0038$                                          & 0.213              \\
		$i$ (deg)       &  $75.46\pm0.75$                                       & 71.5                  \\
		$\Omega$ (deg)  &     $115.9\pm0.72$                                    & 25              \\
		$\omega$ (deg)  & $307.07\pm2.20$                                       & 109.9   \\
		Mass$_{\rm Tot}$  (\rm $M_\odot$) & $2.18\pm0.33$ & 1.03   \\ \hline
	\end{tabular}
\end{table}

We initially obtained large errors from our orbital modeling of the system using the \cite{tokovinin2016orbits} method. This is due to the two points from \cite{1992A&AS...96..375I} and \cite{2000AJ....119.3084H} which deviate significantly from the other orbital points. We therefore assigned less weightage for the two points in our fit, and managed to get the best-fit solution.

We show our orbital solution in Figure~\ref{orbit}, in comparison with that obtained by \cite{docobo2006new}. Based on this figure, it can be seen that the two orbits are significantly different from each other, and our orbit fits with the observed positional measurements better than \cite{docobo2006new}. The average residuals for the fit are -2.77$^\circ$ and 6.26 mas for $\theta$ and $\rho$, respectively, and the root-mean-square (RMS) are 7.8833$^\circ$ and 0.02172 mas for $\theta$ and $\rho$, respectively. We tabulate the orbital properties of the system that we obtained from our analysis in Table~\ref{tab:The-adopted-orbital}.

We can independently calculate the total mass of the system using the orbital properties that we obtained from this dynamical analysis. This can be determined using Kepler's third law, which is given by:
\begin{eqnarray}
\rm Mass_{Tot}=\left (\rm Mass_{A_{a}}+\rm Mass_{A_{b}}+\rm Mass_{B_{a}}+\rm Mass_{B_{b}}\right) /{ M_{\odot}}=(a{{}^3}/\pi{{}^3}P{{}^2})\label{eq:8}
\end{eqnarray}
where the units of $P$ is in years, $a$ and $\pi$ are in arcseconds.
Based on eq. (\ref{eq:8}), we calculated a total dynamical mass of $2.18 \pm 0.33\  M_{\odot}$ for the system. This value is significantly smaller than the total mass determined from our SED analysis; i.e., $3.42\pm 0.20\  M_\odot$. A reason for this could be due to inaccuracy in the parallax measurement. The same mass can be obtained if the parallax of the system was to be $\approx$17.95 mas.

We note that in a private communication with J. A. Docobo during the final submission stage of this paper, we were informed that in their work (Docobo and Campo in preparation), they performed orbital modeling for this system using the same data points as ours, with a different method. The results that they obtained are as follows: $P=20.414$ yr., $T_{\rm o}=1999.240$ yr., $e=0.5238$, $a=0.2011$ arcsec, $i=76.03 ^\circ$, $\Omega=116.69^\circ$ and $\omega=307.21^\circ$, with RMS of 7.636$^\circ$ and 0.0221 mas for $\theta$ and $\rho$ respectively. These orbital elements are highly consistent with that determined in our work (Table 6), providing further support to the results that we obtained.

\section{CONCLUSION}

In this paper, we analyzed the stellar system HIP 101227 to determine the total number of components in the system, and their properties. We used recent data and methods developed by ~\cite{al2002spectral} (atmospheric modeling) and \cite{tokovinin2016orbits} (dynamical modeling) to determine the physical properties and orbital solution for the system, respectively, with better accuracy than past studies. Based on our analysis, we found that the system is more consistent with being a quadruple rather than a binary, or a triple system as that suggested by previous studies. This hierarchical quadruple system consists of four zero-age main sequence stars (at the last stage of the pre-main sequence), with age less than 200 Myr and very similar physical properties. Due to their very similar characteristics, we suggested that fragmentation process as the most probable mechanism for the formation of the system.



\section{ACKNOWLEDGMENTS}

We thank the reviewer for his constructive feedback on our manuscript which have helped to significantly improved the paper. We also thank Andrei Tokovinin for his help in solving the orbit of the system, and for valuable discussion.

A. Annuar acknowledges financial support from Malaysia's Ministry of Higher Education Fundamental Research Grant Scheme code FRGS/1/2019/STG02/UKM/02/7. This work has made use of SAO/NASA, SIMBAD database, Fourth Catalog of Interferometric Measurements of Binary Stars, Sixth Catalog of Orbits of Visual Binary Stars, IPAC data systems,  ORBITX code and Al-Wardat's complex method for analyzing CVBMSs with its codes, written in FORTRAN and  Interactive Data Language (IDL) of the ITT Visual Information Solutions Corporation.

\appendix
\section{Results for binary and triple system analysis}

\begin{figure}[hbt!]
\begin{centering}
\includegraphics[scale=0.55]{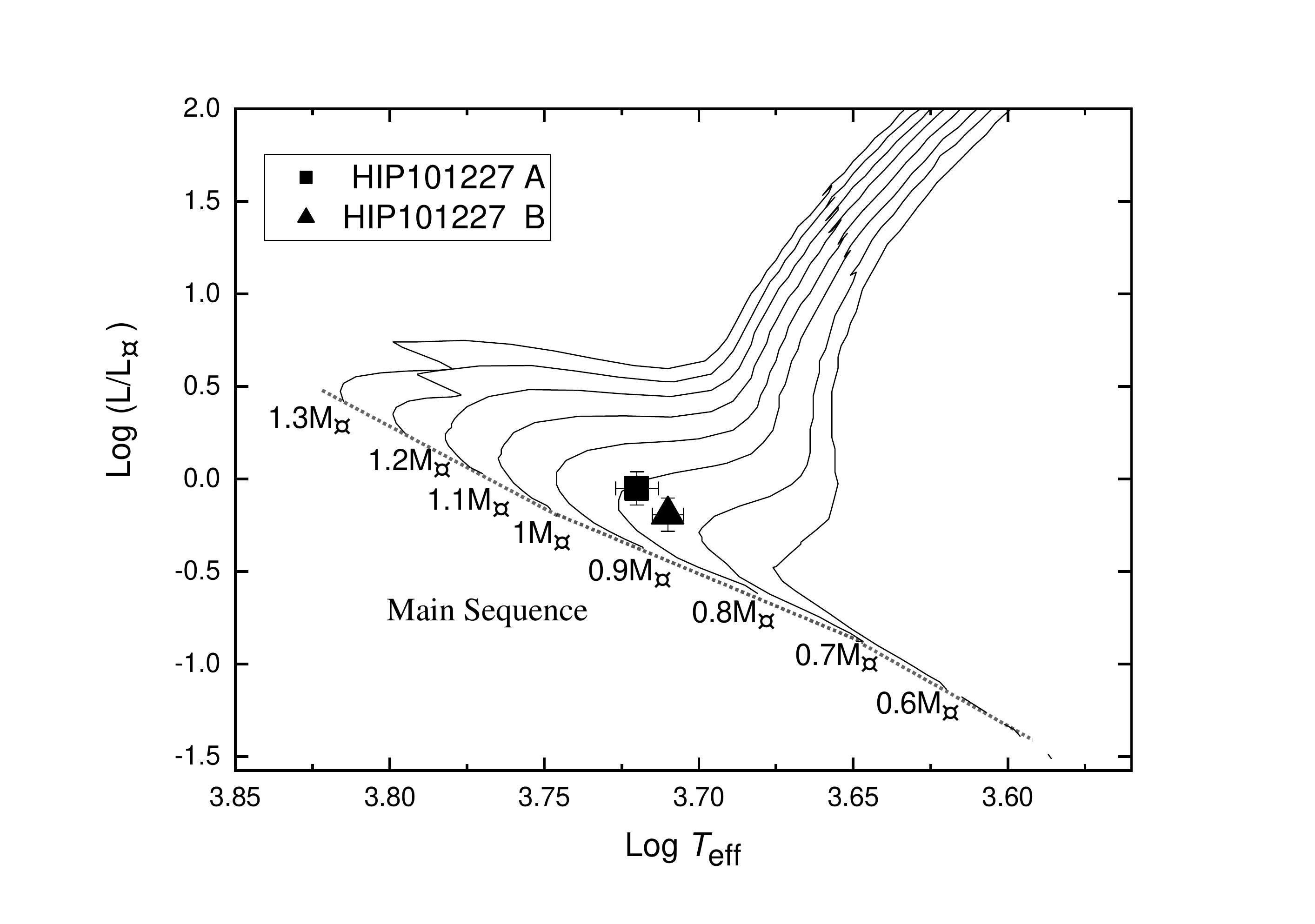}
\caption{The positions of each star in the HIP 101227 system on the evolutionary tracks derived by \cite{girardi2000evolutionary} assuming that it is a binary. \label{fig:evolutionary}}
\par\end{centering}
\end{figure}

\begin{figure}[hbt!]
\begin{centering}
\includegraphics[scale=0.55]{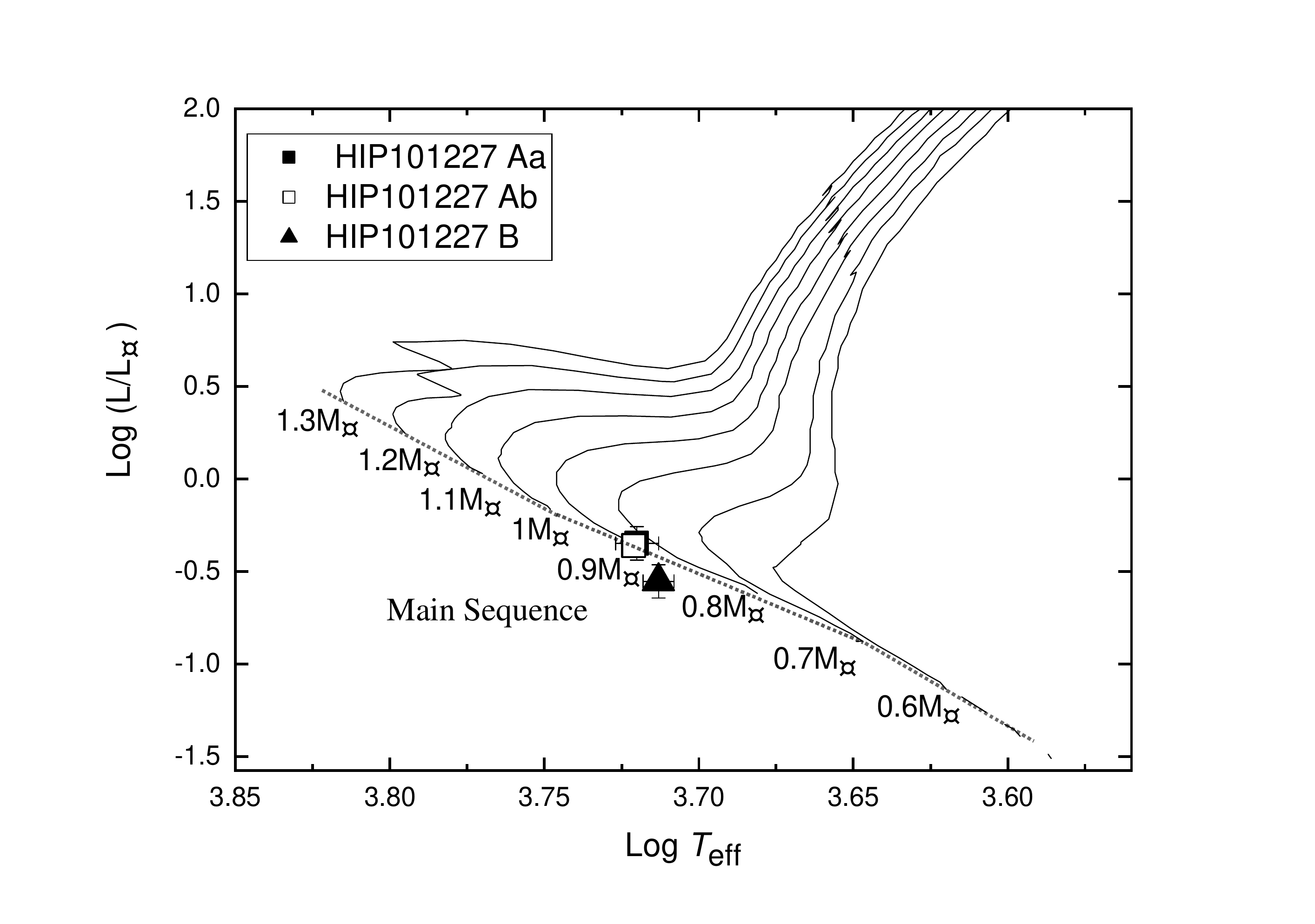}
\includegraphics[scale=0.55]{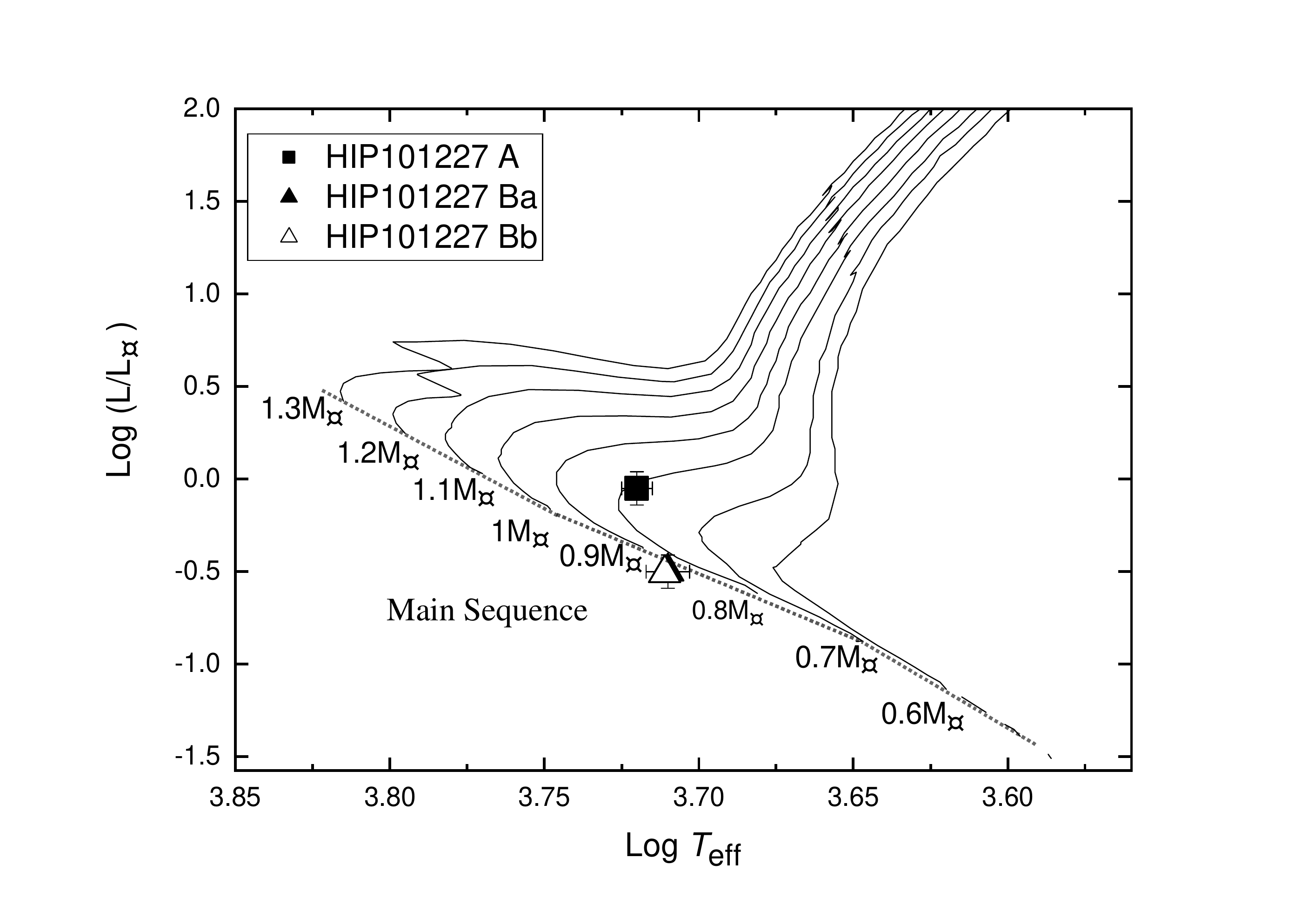}
\caption{The positions of each star in the HIP 101227 system on the evolutionary tracks derived by \cite{girardi2000evolutionary} assuming that it is a triple in A (top) and triple in B (bottom). \label{fig:evolutionarytriple}}
\par\end{centering}
\end{figure}

\begin{figure}[hbt!]
\begin{centering}
\includegraphics[scale=0.55]{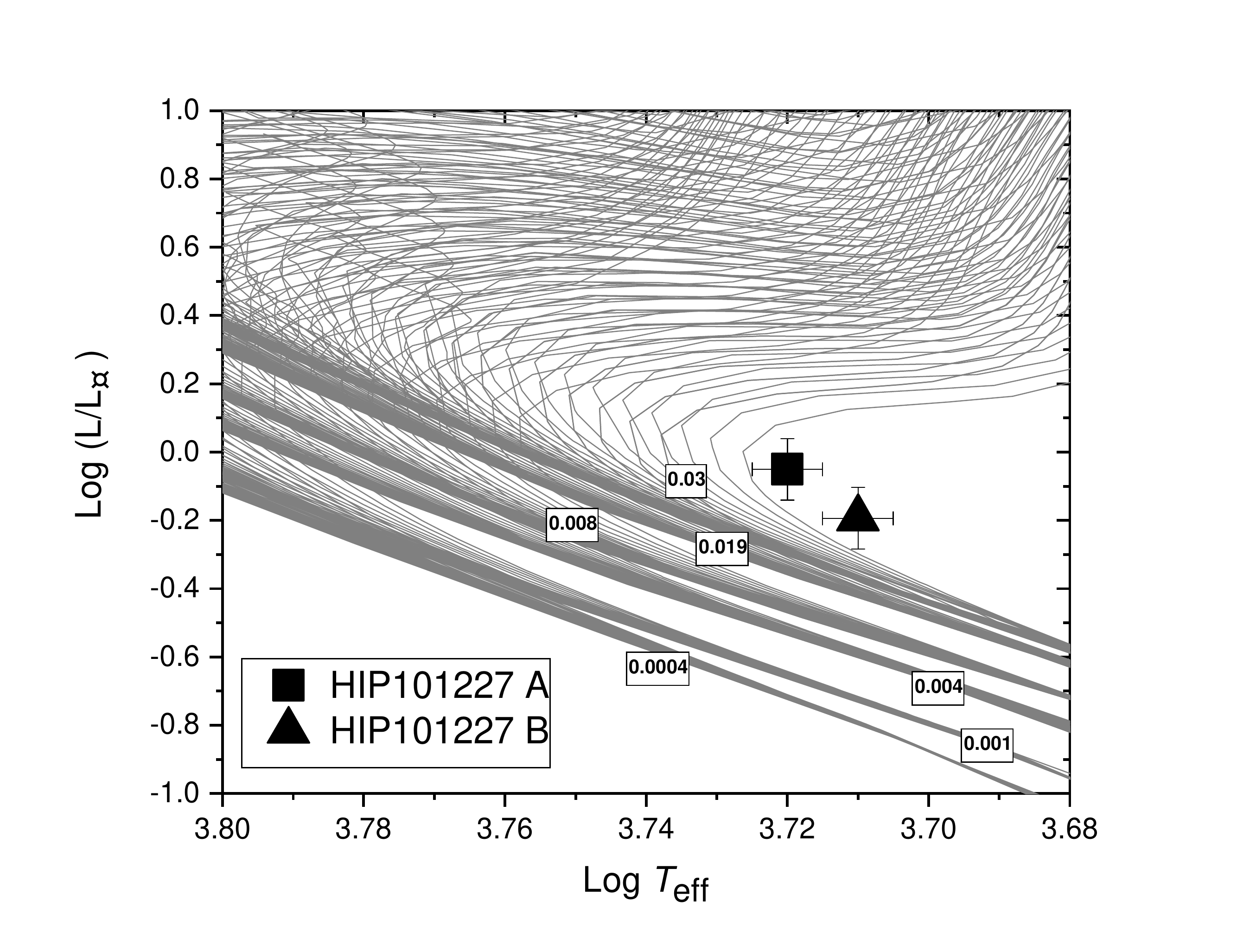}
\caption{The positions of each star in the HIP 101227 system on the isochrones for low and intermediate-mass stars with different metallicities derived by \cite{girardi2000evolutionary} assuming that it is a binary.}
\label{fig:iso-binary}
\par\end{centering}
\end{figure}

\begin{figure}[hbt!]
\begin{centering}
\includegraphics[scale=0.55]{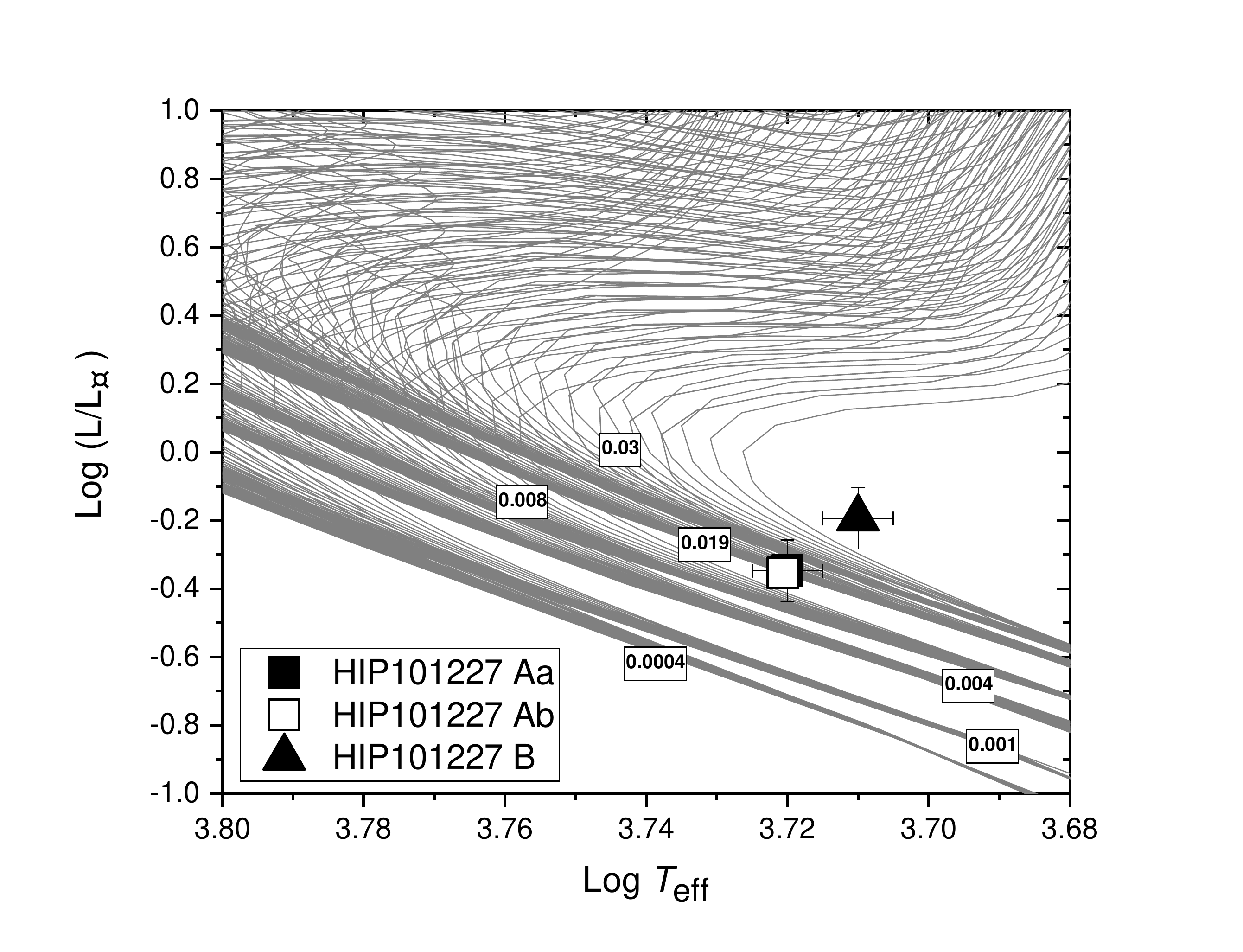}
\includegraphics[scale=0.55]{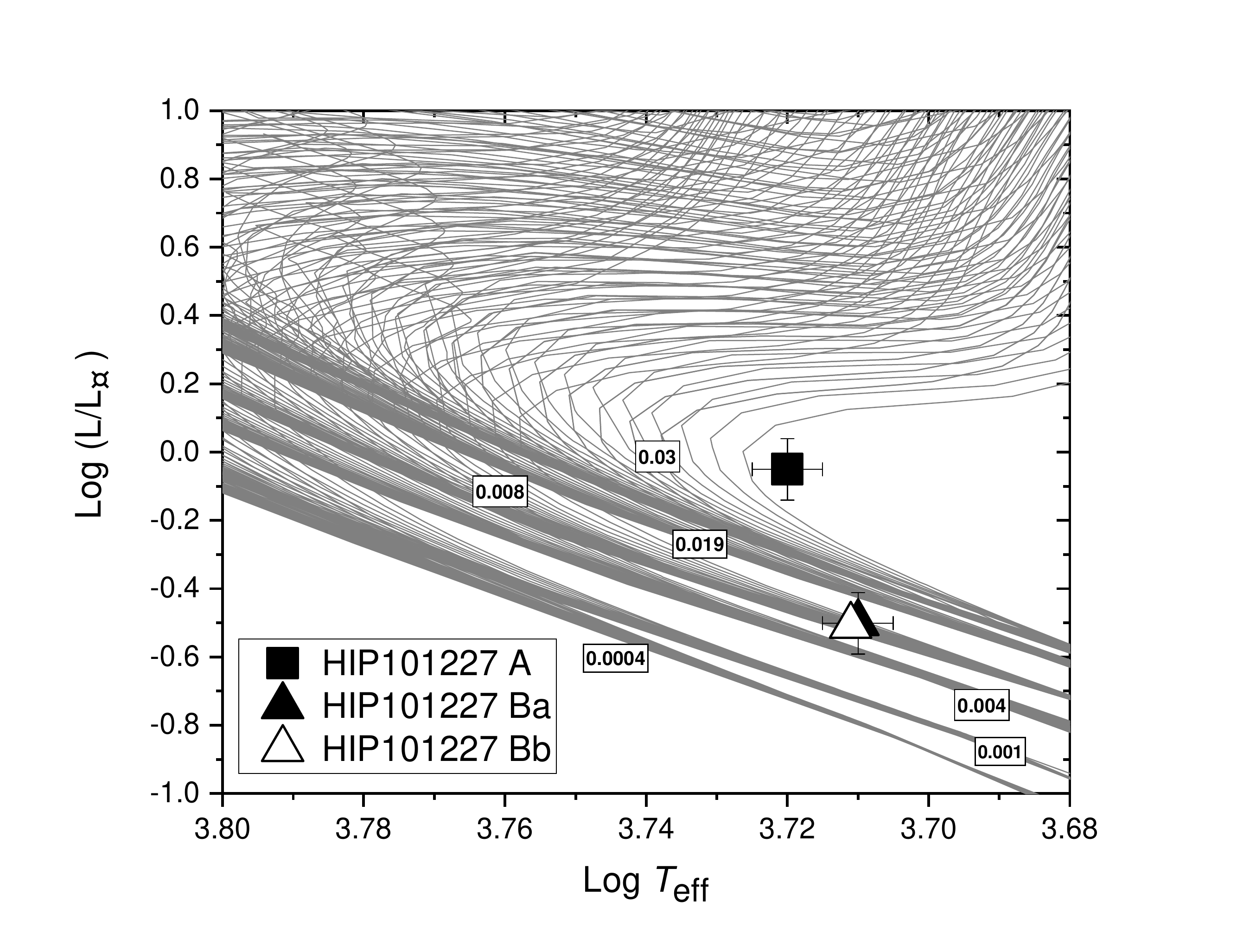}
\caption{The positions of each star in the HIP 101227 system on the isochrones for low and intermediate-mass stars with different metallicities derived by \cite{girardi2000evolutionary} assuming that it is a triple in A (top) and triple in B (bottom).}
\label{fig:iso-triple}
\par\end{centering}
\end{figure}

Here we provide the results of our analysis when assuming that the system is a binary and triple system, which was discussed in Section 3.1. This includes the physical properties that we determined (Table \ref{tab:binary-parameter}-\ref{tab:TRIPLE-B-parameter}), as well as the associated evolutionary track and isochrone diagrams (Figures \ref{fig:evolutionary}-\ref{fig:evolutionarytriple} and Figures \ref{fig:iso-binary}-\ref{fig:iso-triple}, respectively).

\begin{table}[hbt!]
	\caption{Physical properties of each star in the HIP 101227 when assuming that it is a binary system}.\label{tab:binary-parameter}
	\centering{}\begin{tabular}{@{}lcc@{}}
		\hline
	Parameters     & Comp. $A$ & Comp. $B$ 	\\ \hline
$T_{\rm eff}$ (K) & $5250\pm30$   & $5170\pm20$  \\
R ($R_{\odot}$) & $1.14\pm0.01$   & $1.00\pm0.01$   \\
Log $g$   & $4.60\pm0.05$& $4.60\pm0.05$   \\
L ($L_{\odot}$) & $0.89\pm0.11$     & $0.64\pm0.08$  \\
Mass ($M_{\odot}$) & $0.79\pm0.20$     & $0.75\pm0.20$ \\\hline
	\end{tabular}
\end{table}

\begin{table}[hbt!]
	\caption{Physical properties of each star in the HIP 101227 when assuming that it is a triple system, with component $A$ being a sub-binary.\label{tab:triple-A-parameter}}
	\centering{}\begin{tabular}{@{}lccc@{}}
		\hline
	Parameters     & Comp. $A_{\rm a}$ & Comp. $A_{\rm b}$ & Comp. $B$	\\ \hline
$T_{\rm eff} $ (K) & $5250\pm30$    & $5250\pm30$     & $5170\pm20$   \\
R ($R_{\odot}$) & $0.81\pm0.02$     & $0.81\pm0.02$    & $1.00\pm0.01$  \\
 Log $g$  & $4.60\pm0.05$ &  $4.60\pm0.05$ & $4.60\pm0.05$ \\
L ($L_{\odot}$) & $0.45\pm0.05$     & $0.45\pm0.05$   & $ 0.64\pm0.08$  \\
Mass ($M_{\odot}$) & $0.90\pm0.20$     & $0.90\pm0.20$   & $ 0.75\pm0.20$ \\\hline
	\end{tabular}
\end{table}

 \begin{table}[hbt!]
	\caption{Physical properties of each star in the HIP 101227 when assuming that it is a triple system, with component $B$ being a sub-binary.\label{tab:TRIPLE-B-parameter}}
	\centering{}\begin{tabular}{@{}lccc@{}}
		\hline
	Parameters     & Comp. $A$ & Comp. $B_{\rm a}$ & Comp. $B_{\rm b}$ \	\\ \hline
$T_{\rm eff} $ (K)     & $5250\pm30$     & $5170\pm20$   & $5170\pm20$  \\
R ($R_{\odot}$) & $1.14\pm0.01$    &$0.70\pm0.01$   & $0.70\pm0.01$ \\

 Log $g$   & $4.60\pm0.05$& $4.60\pm0.05$ & $4.60\pm0.05$\\
L ($L_{\odot}$)  & $0.89\pm0.11$   & $ 0.31\pm0.05$   & $ 0.31\pm0.05$  \\
Mass ($M_{\odot}$) & $0.79\pm0.20$  & $ 0.81\pm0.20$   & $ 0.81\pm0.20$ \\\hline
	\end{tabular}
\end{table}

\clearpage


\end{document}